\documentstyle[twoside,fleqn,npb,epsfig]{article}

\newcommand{\AmS}{{\protect\the\textfont2
  A\kern-.1667em\lower.5ex\hbox{M}\kern-.125emS}}

\title{Renormalization group improved BFKL equation
	\thanks{Work supported in part by the E.U. QCDNET contract FMRX-CT98-0194 
	and by MURST (Italy)}
	}

\author{Marcello Ciafaloni
	\address{Dipartimento di Fisica dell'Universit\`a, 
	Firenze \\
        and INFN, Sezione di Firenze, Italy}%
        }

\begin{document}

\begin{abstract}
I report on the recent proposal of a generalized small-$x$ equation which,
in addition to exact leading and next-to-leading BFKL kernels, incorporates
renormalization group constraints in the relevant collinear limits.
\end{abstract}

\maketitle

The calculation of next-to-leading log $x$ corrections to the BFKL
equation was completed last year \cite{fadin,camici} after several
years of theoretical effort. The results, for both anomalous dimension
and hard Pomeron, show however signs of instability due to both the
size and the (negative) sign of corrections, possibly leading to 
problems with positivity also \cite{ross}.

If we write the eigenvalue equation corresponding to the BFKL solution
in the form \cite{camici}
\begin{eqnarray}
\omega &=& \bar{\alpha}_{s} (t) \left[ \chi_{0} (\gamma) +
\bar{\alpha}_{s} (\mu^{2}) \chi_{1} (\gamma) + \dots \right], 
\nonumber \\
t &=& \log \frac{k^{2}}{\lambda^{2}} ,
\label{1}
\end{eqnarray}
where $\omega = N-1$ is the moment index and $\gamma$ is an
anomalous dimension variable, the NL eigenvalue function  has the shape of
Fig 1, which completely overtrows the LL picture, even for
coupling values as low as .04.

The basic reason for the instability above lies in the $\gamma$-singularity
structure of $\chi_{1}$ (cubic poles) which are of collinear origin,
and keep track of the choice of the scaling variable, whether it is
$k k_{0}$/s, or $k^{2}$/s, or $k^{2}_{0}$/s, in a two-scale hard
process. An additional reason lies in the renormalization scale
$(\mu)$ dependence of Eq (1), related to the method of solution.

In a recent proposal \cite{salam,ciafa2}, both problems are
overcome at once by a proper use of R.G. constraints on both
kernel and solution. On one hand, the requirement of single-log
scaling violations for both $k \gg k_{0}$ with Bjorken variable
$k^{2}$/s and for the symmetrical limit, imply an $\omega$-dependent
shift of the $\gamma$-singularities in the kernel which resums
the double logarithmic ones mentioned before.
\begin{figure}[hp]
\begin{picture}(70,80)(0,0)
\epsfig{file=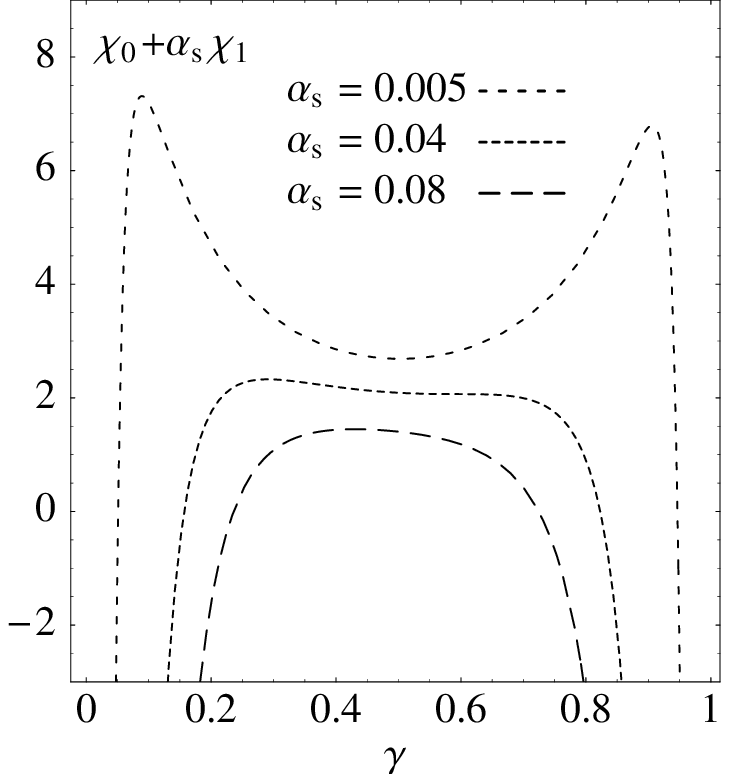,width=75mm}
\end{picture}
\caption{The BFKL eigenvalue function at NL accuracy,
              for $\alpha_{s}$ = 0.005, 0.04 and 0.08.}
\end{figure}

On the other hand, a novel method of solution called $\omega$-expansion
\cite{ciafa1} replaces $\alpha_{s}$ with $\omega$ as perturbative
parameter of the subleading hierarchy, and allows a R.G. invariant
formulation of the solution. More precisely, for large $t$ the gluon
Green's function takes the factorized form
\begin{equation}
G_{\omega} ({\mathbf k}, {\mathbf k}_{0} )  = {\cal F}_\omega
({\mathbf k}) \tilde{\cal F}_{\omega} ({\mathbf k}_{0}), \,
t-t_{0} \gg 1 ,
\label{2}
\end{equation}
where
\begin{eqnarray}
\dot{g}_{\omega} (t) &\sim& {\mathbf k}^{2} {\cal F}_{\omega} ({\mathbf k})
\nonumber \\
&=& \int \frac{d\gamma}{2 \pi i} \exp \left[ 
\gamma t - \frac{1}{b\omega} X (\gamma , \omega) \right]
\label{3}
\end{eqnarray}
is the $t$-dependent unintegrated gluon density. The phase function
X is given in terms of the effective eigenvalue function
\begin{equation}
\frac{\partial}{\partial\gamma} X (\gamma , \omega) = \chi
(\gamma , \omega) = \chi^{\omega}_{0} (\gamma) +
\omega \frac{\chi^{\omega}_{1} (\gamma)}{\chi^{\omega}_{0} (\gamma)}
+ \dots ,
\label{4}
\end{equation}
which now has a fully stable $\omega$-dependence (Fig.2).

\begin{figure}[hp]
\begin{picture}(75,80)(0,0)
\epsfig{file=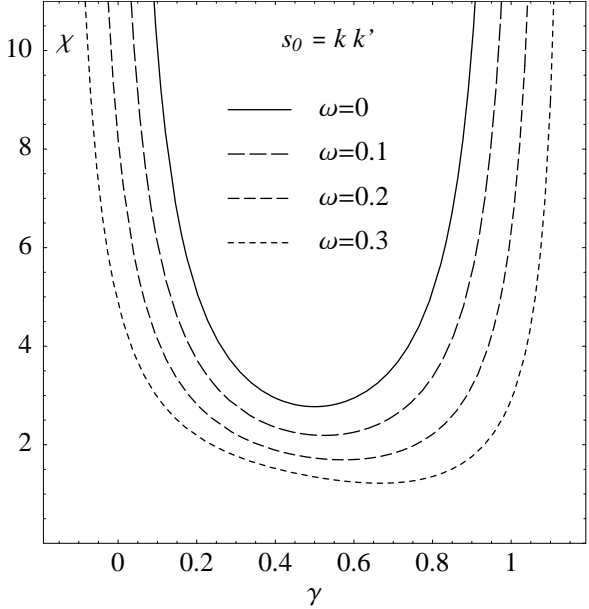,width=75mm}
\end{picture}
\caption{The effective eigenvalue function $\chi(\gamma,\omega)$
for various $\omega$ values.}
\label{fig:2}
\end{figure}

The improved kernel eigenvalue functions $\chi^{\omega}_{0}$ and
$\chi^{\omega}_{1}$ are constructed from the exact L+NL kernels,
by incorporating the $\omega$-shift requirement. The neglected
terms in Eq.4 yield a small error, corresponding to a coupling
change $\delta \alpha_{s}/\alpha_{s} = O (\alpha_{s} \omega)$,
subleading in both $\frac{\alpha_{s}}{\omega}$ and $\alpha_{s}$
expansions of the small-$x$ hierarchy.
\begin{figure}[hp]
\begin{picture}(75,72)(0,0)
\epsfig{file=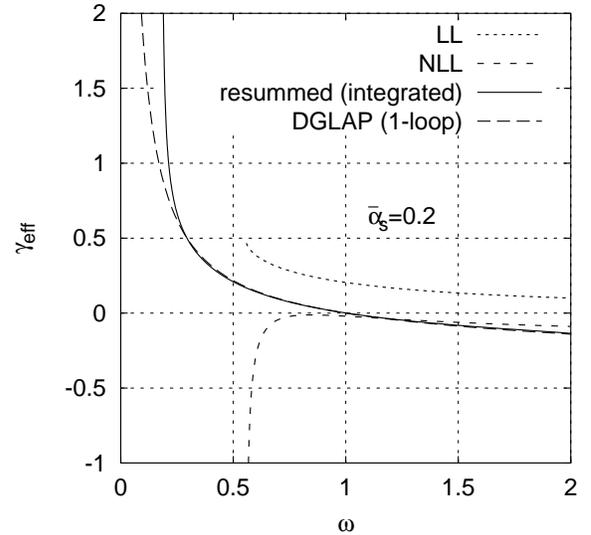,width=75mm}
\end{picture}
\caption{The resummed gluon anomalous dimension compared
to various approximations.}
\label{fig:3}
\end{figure}

The solution for the effective anomalous dimension $\gamma_{eff}
= \dot{g}_{\omega} (t) /g_{\omega} (t)$ is shown in Fig.3,
compared to L and NL approximations. The resummed result is
remarkably similar to the fixed order value until very close
to the singularity point $\omega_{c}(t)$, which lies below the
saddle point breakdown value $\omega_{s}(t)$ used in previous
NL estimates of the hard Pomeron. The latter signals the failure
of the large-$t$ saddle point $b\omega t = \chi (\bar{\gamma} ,
\omega) $ to yield a reliable anomalous dimension $\bar{\gamma}$,
due to infinite $\gamma$-fluctuations. The former is the position
of the true $t$-dependent $\omega$ singularity, and is
systematically lower [Fig.4]. No instabilities and very little
renormalization scheme dependence are found.

The critical exponents $\omega_{c} (t)$ and $\omega_{s} (t)$ are 
actually both needed for a full understanding of the Green's
function (2), whose coefficient $\tilde{\cal F}_{\omega} (\mathbf{k}_{0})$
carries the $t$-independent, leading Pomeron singularity, which
is really nonperturbative. While a precise estimate of the latter
requires extrapolating the small-$x$ equation in the strong-coupling
region $k^{2} \simeq \Lambda ^{2}$, one can argue \cite{ciafa2}
that $\omega_{c}$ and $\omega_{s}$ provide lower and upper
bounds on $\omega_{P}$, and thus a first rough estimate of the 
Pomeron intercept.
\begin{figure}[hp]
\begin{picture}(75,80)(0,0)
\epsfig{file=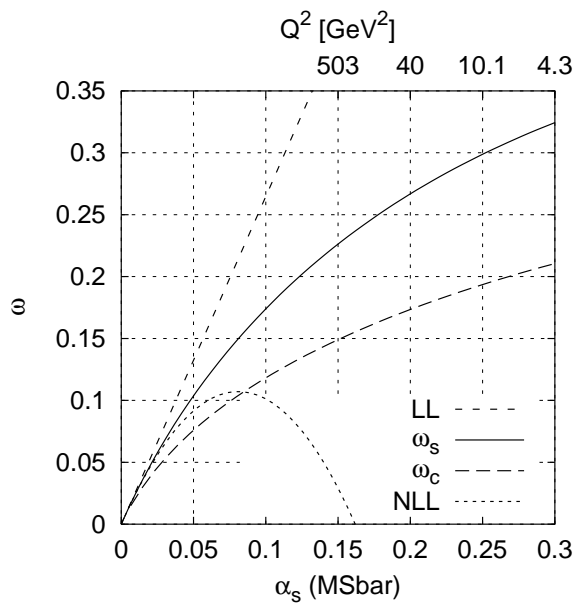,width=75mm}
\end{picture}
\caption{The resummed critical exponents $\omega_{c}(t)$ and
$\omega_{s}(t)$, compared to L and NL estimates of the hard
Pomeron.}
\label{fig:4}
\end{figure}

\vspace{1cm}

I wish to thank Dimitri Colferai  and Gavin Salam for friendly and helpful
discussions.

\end{document}